\documentclass[aip,reprint,graphicx]{revtex4-1} 

\usepackage[english]{babel}
\usepackage[alsoload=synchem, separate-uncertainty=true, multi-part-units=repeat]{siunitx} 
\usepackage[pdftex]{graphicx} 
\usepackage{amsmath}
\usepackage{subcaption}

\draft 

\begin{document}
    

\title{A System for Trapping Barium Ions in a Microfabricated Surface Trap} 




\author{R.D. Graham} \email[]{rdgraham@uw.edu}
\affiliation{University of Washington, Department of Physics, Box 351560, Seattle, WA 98195-1560, USA}

\author{S.-P.\ Chen}
\affiliation{University of Washington, Department of Electrical Engineering, 185 Stevens Way, Paul Allen Center - Room AE100R, Campus Box 352500, Seattle, WA 98195-2500, USA
}

\author{T.\ Sakrejda}
\author{J.\ Wright}
\author{Z.\ Zhou}
\author{B.B.\ Blinov} \email[]{blinov@uw.edu}
\affiliation{University of Washington, Department of Physics, Box 351560, Seattle, WA 98195-1560, USA}


\date{\today}

\begin{abstract}
	    \noindent We have developed a vacuum chamber and control system for rapid testing of microfabricated surface ion traps. Our system is modular in design and is based on an in-vacuum printed circuit board with integrated filters. We have used this system to successfully trap and cool barium ions and have achieved ion `dark' lifetimes of \SI{31.6 \pm 3.4}{\second} with controlled shuttling of ions. We provide a detailed description of the ion trap system including the in-vacuum materials used, control electronics and neutral atom source. We discuss the challenges presented in achieving a system which can work reliably over two years of operations in which the trap under test was changed at least 10 times.
\end{abstract}


\maketitle 

    Trapped ions are a promising candidate building block for a quantum computer. One of the most promising trap designs is the planar ion trap. Here, RF rails and DC bias electrodes are fabricated on an aluminum surface (sometimes gold coated) on silicon using semiconductor fabrication techniques \cite{Seidelin.prl.2006, stick.nature.2005}. The basic design of these traps puts an ion in a RF pseudopotential null \SIrange{50}{200}{\micro\meter} above the chip surface. A vapor of neutral atoms flows through an aperture in the back of the chip. Photo-ionization and laser cooling is done with beams passing just above the surface. After loading ions, the position of the trapping region can be finely controlled by adjusting the DC voltages on multiple electrodes on the trap surface while moving the lasers to follow the ion. Ultimately multiple trapping regions can be defined and ion chains can be re-organized by shuttling ions to different regions.
    
    Chips with several designs of trapping regions have been produced. These include linear traps \cite{Amini.gtriposter.2011, allcock.apb.2012}, multiple arm `Y' \cite{amini.njp.2010, moehring.njp.2011} and ring traps \cite{tabakov.baps.2012}. Novel features of recently fabricated chips include integrated optics for fluorescence light collection \cite{merrill.njp.2011, brady.apb.2011}, integrated photodetectors \cite{eltony.aip.2013}, integrated light sources \cite{kim.aip.2011} and integrated microwave waveguides \cite{allcock.aip.2013}.

    Chip traps are currently fabricated by several groups around the world. We are working with devices provided by Sandia National Labs and Georgia Tech Research Institute (GTRI). Both these groups have standardized on a 100-pin architecture which gives the ability to control up to 96 DC electrodes. Generating all the necessary control voltages and connecting them through an ultra-high vacuum (UHV) interface in a reliable and low noise way is significantly more challenging than in a traditional macroscopic ion trap.
    
    $^{138}$Ba+ is an alkaline rare-earth metal isotope with nuclear spin-0. A key advantage of barium is that all important transitions for cooling and photo-ionization reside within either the visible or near-infrared spectrum, where coherent light sources are easily obtained and transmitted with optical fibers. This feature makes barium a good candidate for implementing a photonic link qubit as part of a large scale quantum computing architecture. Barium ions have been trapped in macroscopic traps of various designs since the 1980's \cite{neuhauser.pra.1980.firstba, Sauter.prl.1986, steele.prb.2007.photoionization, Shu.TackTrap.2011}, though trapping of barium in a chip trap has not been demonstrated before.

    \section{Chamber Design}

    Traditionally building an ion trap in a vacuum system involves wire wrapping and spot welding to make electrical connections. These techniques are well known to be compatible with the UHV environment and continued to be used once chip traps were developed. However, it was soon realized that these techniques do not scale well to chip traps which might require up to 100 connections to be made. Trap assembly is made tedious and error prone. Additionally, large numbers of point to point connections can reduce optical access and impede molecular flow.
    
    In order to facilitate easy mounting and frequent upgrades we have designed a UHV-compatible printed circuit board (PCB) that holds a custom Zero-Insertion-Force (ZIF) socket\footnote{Tactic electronics part 100-4680-001A} for the chip trap. This socket accepts standard 100-pin ceramic Pin Grid Array (PGA) chip carriers used by GTRI and Sandia National Labs. These are carriers are of a square design, 33.5 mm on each side with 0.5 mm pins arranged on the outside.
    
    The circuit board plugs in to a set of four DB-25 female/female adapters\footnote{Accu-Glass Products part 104101} which in turn are connected to feedthroughs integrated into the bottom flange\footnote{Accu-Glass Products part 25D4-133-CF600TAPPED} of the vacuum chamber. A grounding shield with a window of thin wire mesh clips on to the top of the chip carrier. The circuit board incorporates \SI{20}{\nano\farad} 0805 series all-ceramic capacitors approximately 1 cm from the chip for filtering RF pickup on the DC control pins.
    
    The circuit board is made of ceramic-filled Polytetrafluoroethylene (PTFE) material\footnote{Rogers ceramic (RT/duroid\textsuperscript{\textregistered} 6002)} of \SI{1.52}{\milli\meter} thickness with a \SI{35}{\micro\meter} copper cladding. It is a double sided board with plated through-holes. A particular difficulty with this material is its softness which causes it to be easily deformed by insertion pressure. In earlier iterations of the design without a ZIF socket these deformations lead to hairline cracks in the traces under the chip.
    
    Both the ZIF socket, DB-25 adapters and a small spacer on which the ZIF socket sits\footnote{Manufactured in-house} are made of Polyether Ether Ketone (PEEK). The receptacle pins for holding the ZIF socket are a gold plated brass alloy\footnote{Mill-Max part 0326-3-19-15-06-27-10-0}, as are the pins which plug into the DB-25 connectors\footnote{Keystone Electronics part 1358-2}. The pins for connecting the RF rails are longer\footnote{Mill-Max part 0038-3-17-15-30-27-02-0} and pass through the circuit board without connection. They are then wire-wrapped\footnote{30 AWG, silver plated copper wire with Kapton shielding} to a secondary feedthrough on the vacuum chamber. All on-board connections are soldered\footnote{Kester no-clean lead free solder (95.5\% Sn, 3\% Ag, 0.5\% Cu)} and ultrasonically cleaned with flux remover\footnote{Chemtronics ES132} followed by acetone before assembly.
    
\begin{figure}
        \centering
        \includegraphics[width=0.45\textwidth]{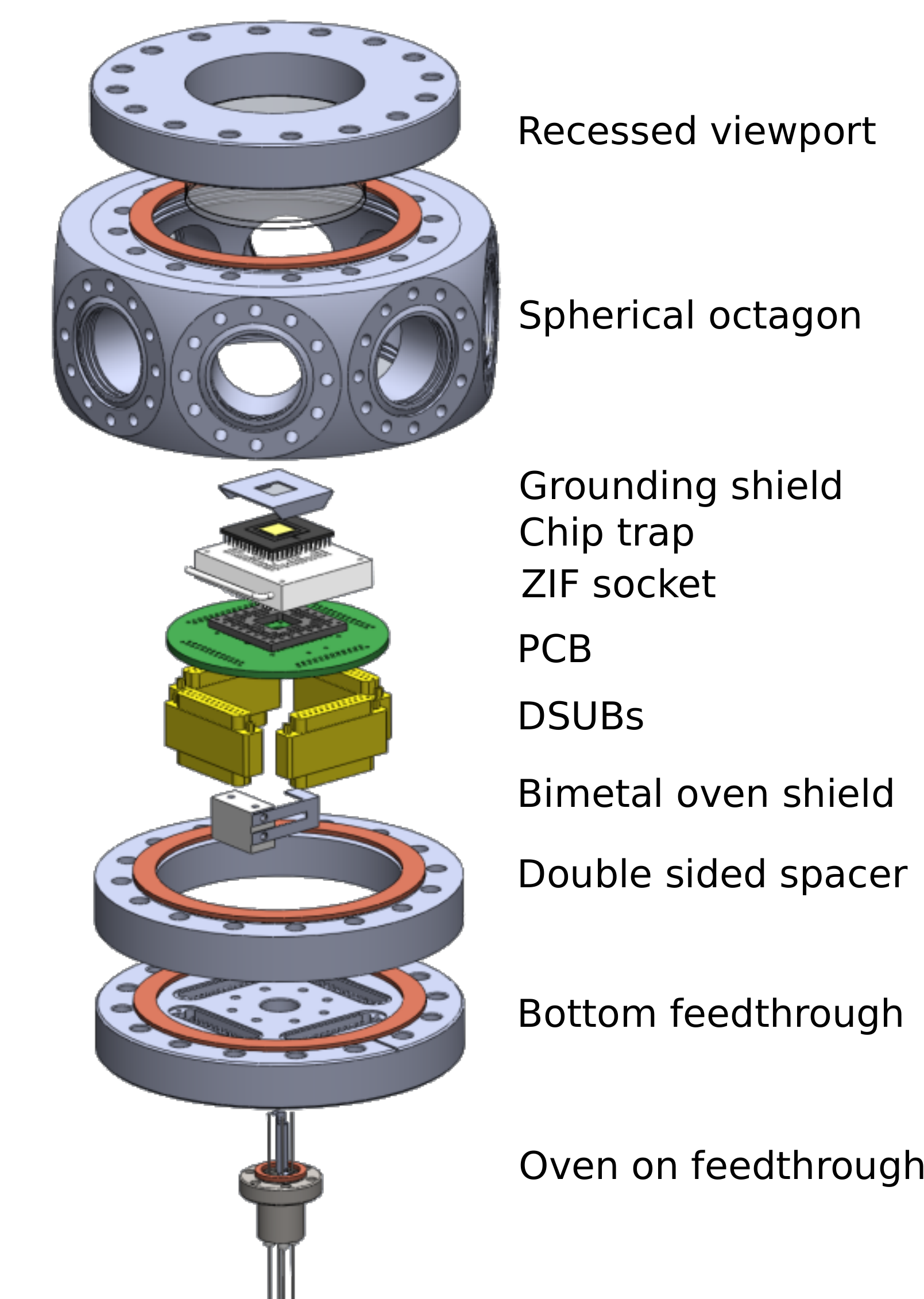}
        \caption{Exploded view of trap chamber showing how the PCB, ZIF socket and chip trap are assembled in a spherical octagon UHV chamber.}
        \label{fig:chamber}
    \end{figure}

    The chamber is a `spherical octagon' design\footnote{Kimbal Physics part MCF600-SphOct-F2C8}. The chamber has eight 2.75 inch side ports, six of which provide optical access for the laser beams. Optical access for imaging occurs through a viewport on the top 6 inch port. Vertical space of approximately \SI{25}{\mm} is available underneath the circuit board for mounting the atomic vapor source. Figure~\ref{fig:chamber} shows an exploded view of the trap chamber.
    
    Pumping is done with an on-system 20~L/s ion pump and a titanium sublimation pump, each connected to a 2.75 inch 5-way junction on one of the chamber ports. After baking at \SI{150}{\degreeCelsius} for 1--2 weeks we achieve a base pressure of a few $10^{-11}$~Torr.
    
    \section{Optical setup}
    
    The first step required in loading ions from an atomic vapor source is photo-ionization, either with a single UV source or combination of sources. Direct barium photo-ionization requires greater than \SI{5.21}{\eV} photon energy (\SI{238}{\nm}). The accumulation of UV induced charges on trap components such as Macor\textsuperscript{\textregistered} parts is a known problem with wavelengths below \SI{500}{\nm} and increases rapidly with shorter wavelengths \cite{harter.1305.6826}. The effect on PEEK and ceramic filled PTFE has not been characterized. In a chip trap surface charging is even more of a concern because the lasers must pass just 70 um above the whole surface of the chip (around 1 cm path length). Therefore we require well focused beams with small divergence and wavelength as long as practical.
    
    In barium, the traditional approach to photo-ionization has been to use a \SI{791}{\nm} laser to excite to the $6s6p \: ^{3}P_{1}$ state (\SI{1.57}{\eV}), followed by a UV flash from a \SI{337}{\nm} nitrogen laser (\SI{3.68}{\eV}) or UV flash lamp to complete the ionization \cite{steele.prb.2007.photoionization}. An alternative approaches uses a 2-photon scheme with a single laser at \SI{413}{\nm} initially exciting the $5d6p \: ^{3}D_{1}$ and then ionizing \cite{rotter.2008}. Another recently demonstrated alternative uses a \SI{553}{\nm} laser to excite the strong dipole transition $6s^2 \: ^1S_0 \rightarrow 6s6p \: ^1P_1$ followed by any laser with wavelength less than \SI{418}{\nm} to ionize \cite{Schaetz.apb.2012}.
    
    To eliminate the use of far UV sources we use an ionization scheme in which we retain the traditional \SI{791}{\nm} laser and employ a secondary transition, $6s6p \rightarrow 6p^2$ (\SI{2.75}{\eV}) driven with a \SI{450}{\nm} laser\footnote{In house design ECDL}. Finally, a tertiary transition is required, the \SI{791}{\nm} laser (or other cooling laser) can provide the necessary \SI{.91}{\eV} to complete the ionization. An additional advantage of this scheme is improved isotope selectivity.
    
    One difficulty encountered with this scheme is that after loading we observe immediate `shelving' of the ion into the $D_{5/2}$ state. This process can be explained by a transition via the $7S_{1/2}$ level in the barium ion driven by incoherent light from the \SI{450}{\nm} laser. To solve this we require a third laser for de-shelving at \SI{614}{\nm}. This laser is obtained by frequency doubling a \SI{1228}{\nm} ECDL\footnote{Toptica DL-Pro}. The addition of this laser is not an inconvenience because it can serve as a de-shelving laser for $D_{5/2}$ often needed in experiments.
        
    Primary doppler cooling is done on the $S_{1/2}\rightarrow P_{1/2}$ transition at \SI{493}{\nm}. This beam is derived by frequency doubling\footnote{Custom frequency doubling crystal from HCP photonics, single fiber pigtail.} a \SI{986}{\nm} external cavity diode laser\footnote{Toptica DL-100} (ECDL). This doubling arrangement is particularly advantageous because it allows fast switching via pickoff of a first order beam from an acousto-optic modulator (AOM) to be done in the IR. The isolation ratio when switching has traditionally been a problem in experiments. Due to the non-linearity of second-harmonic generation we achieve a very high isolation ratio which is not directly measurable. The lower limit to our isolation ratio measurement is \SI{43}{\decibel}.
    
    A \SI{650}{\nm} laser is also required to re-pump out of the $D_{3/2}$ state. This is obtained from another ECDL\footnote{Toptica DL-100}. Detecting the presence of poorly cooled ions in the initial stages of trapping is done by momentarily switching off this \SI{650}{\nm} laser and observing a decrease in \SI{493}{\nm} fluorescence signal.
    
\begin{figure}
        \centering
        \includegraphics[width=0.45\textwidth]{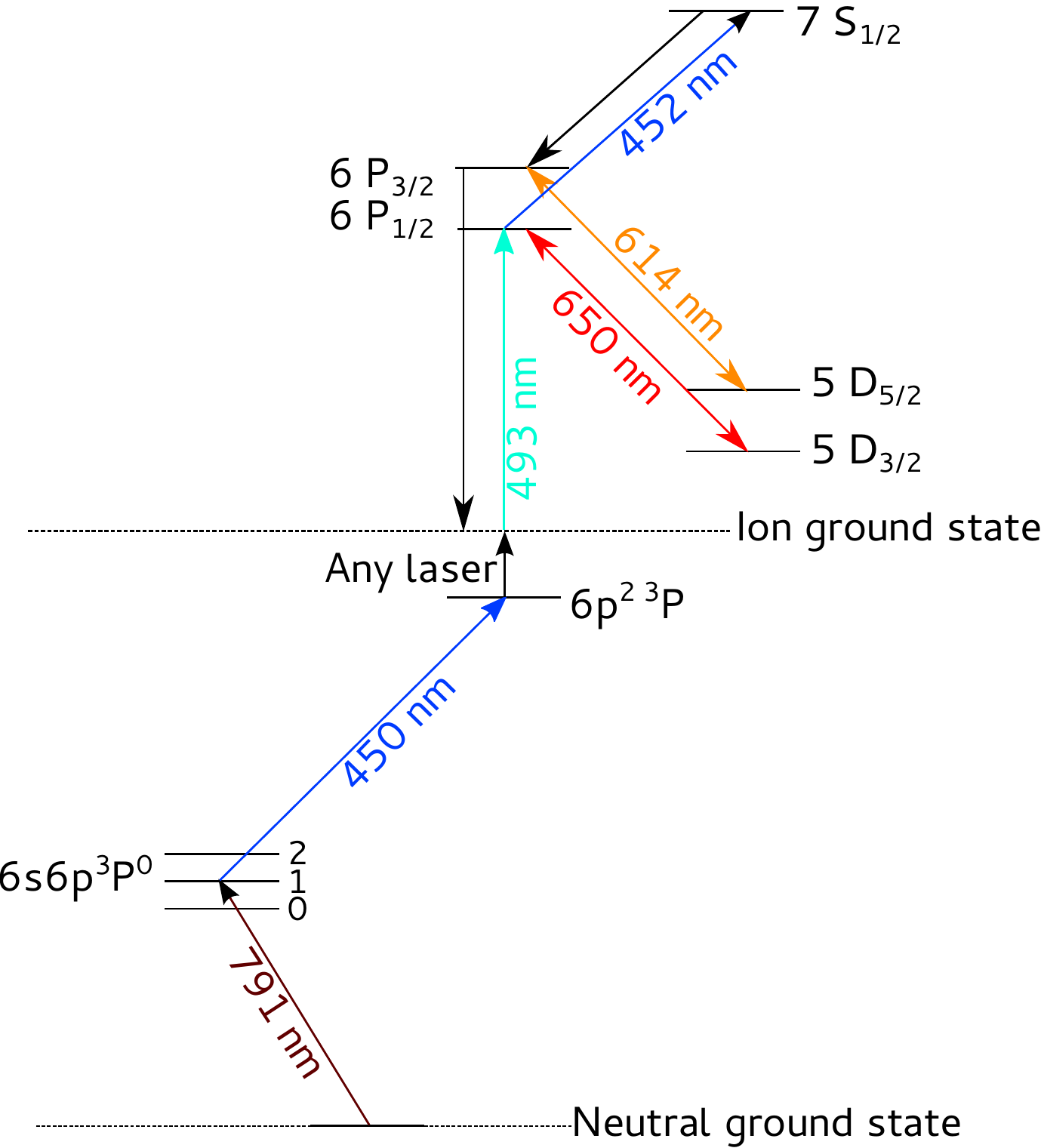}
        \caption{Energy levels and laser transitions relevant for ionization and ion cooling in $^{138}$Ba. Level information from \cite{ba-ion-transitions, ba-all-transitions}. Approximate wavelengths of the laser driven transitions show, for exact frequencies see Table~\ref{tab:lasers}.}
        \label{fig:Ba-levels}
    \end{figure}

    An energy level diagram showing the relevant transitions in neutral and ionized $^{138}$Ba is shown in Figure~\ref{fig:Ba-levels}. Further details of the barium doppler cooling have been presented by others\cite{nagourney.prl.1986.shelved, steele.prb.2007.photoionization, DeVoe.pra.2002.heating}.
    
\begin{figure*}
        \centering
        \includegraphics[width=.75\textwidth]{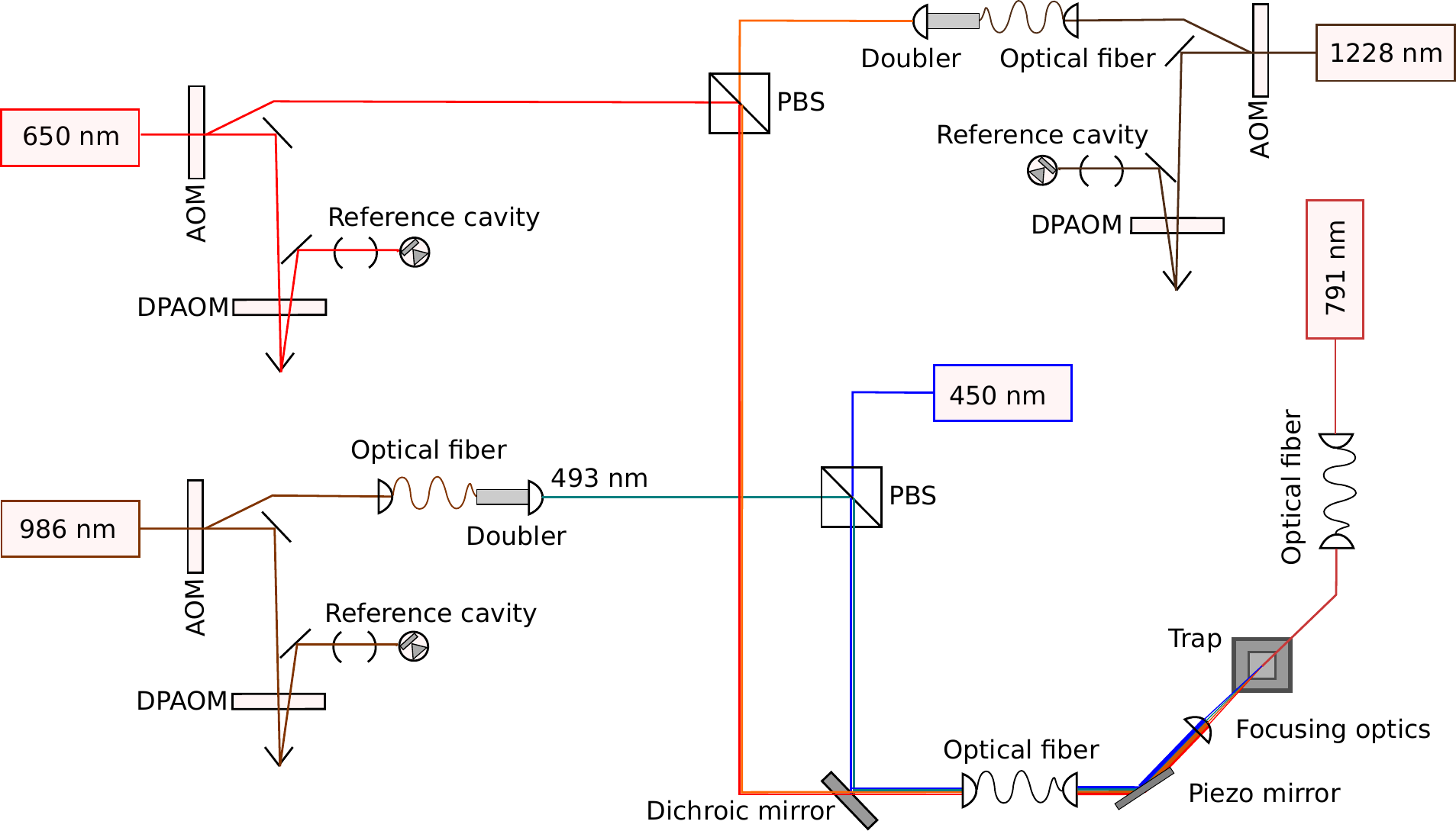}
        \caption{Simplified schematic of the optical layout of the lasers required for ionization and laser cooling of barium. Note that the 4 visible lasers are coupled into a single optical fiber which greatly simplifies the delivery of light to the ion. Not shown are pickoffs for each of the beams to a wavelength meter which is used for tuning the exact laser frequencies.}
        \label{fig:layout}
    \end{figure*}

    Figure~\ref{fig:layout} shows a simplified representation of the optical setup for a single beamline. The \SI{986}{\nm}, \SI{650}{\nm} and \SI{1228}{\nm} lasers are each frequency locked to temperature stabilized invar reference cavities in a hermetically sealed chamber using a custom top-of-the-fringe locking system. In each a zeroth-order beam from a switching AOM is passed through a double-passed AOM (DPAOM) setup to derive the beam to scan the cavity.
    
    All of the beams except the \SI{791}{\nm} beam reside within the transmission spectrum of Nufern 460-HP optical fiber\footnote{Strictly \SI{650}{\nm} is outside the transmission range but in practice can be coupled without difficulty in some fibers of this type due to manufacturing variations.} and are coupled into a single optical fiber. This is particularly convenient for trapping operations because all four beams can be focused on the ion with a single setup. Coupling is achieved by first arranging combined `blue' and `red' beams from the \SI{493}{\nm} + \SI{450}{\nm} and \SI{650}{\nm} + \SI{614}{\nm} beams respectively each using polarizing beam splitters. These two combined beams are then further combined using a dichroic mirror\footnote{Thor labs DMLP567} and coupled into the angle-cleaved end of an optical fiber.
    
    The output of the fiber has a plane-cleaved end and an achromatic microscope objective is used to collimate the output beams. This is reflected by a piezo-controlled mirror\footnote{Custom design.} which allows the beam to be automatically pointed to the new ion position after shuttling. Finally a translatable lens is used for fine adjustment to focus the beams at the trapping region. Typical power levels for each of the beams at the fiber output and nominal trapping frequencies are shown in Table~\ref{tab:lasers}.
    
    Efficient collection of fluorescence light from the ion is important for robust state detection. Our chamber setup makes use of a custom top mounted recessed fused silica viewport which allows imaging optics to be placed within \SI{20.7}{\mm} of the trap surface. Otherwise our imaging system is conventional, employing a 20x microscope objective\footnote{Mittitoyo M-Plan APO} and second stage doublet lens.
    
    The imaging beam path is split by a 50/50 beam splitter between a photomultiplier tube\footnote{Hammamatsu H10682-210} (PMT) configured for photon counting and an electron-multiplying CCD camera\footnote{Luca R series from Andor Technology}. A \SI{493}{\nm} optical bandpass filter can be switched in to the imaging beam path to block background light. Typical PMT count rates observed are \num{3000}~counts/sec/ion against a background of \num{500}~counts/s. Achieving a low background count rate requires tight focusing of beams and precise leveling relative to the chip surface.
    
    The imaging stack is mounted on a translation stage which is actuated by NEMA-23 stepper motors coupled to the micrometers via a helical coupler and brass sliding square peg coupler. This allows the imaging system to be moved automatically to image ions in different locations on the trap surface.
    
    \begin{table*}
        \centering{}
        \begin{tabular}{ccS[table-format=3.5]c}
            \toprule
            Laser  & Approx wavelength & \multicolumn{1}{c}{Exact frequency} & Power at ion \\
                   & nm                & \multicolumn{1}{c}{THz}             & $\mu$W \\
            \colrule
            Primary ionization   & 791  & 378.83665                        & 250 \\
            De-shelving          & 614  & 487.99008                        & 0.5 \\
            Re-pump              & 650  & 461.31197                        & 50 \\
            Secondary ionization & 450  & 665.1433                         & 250 \\
            Cooling              & 493  & 607.42659                        & 5 \\
            \botrule
        \end{tabular}
        \caption{Nominal laser operating frequencies and powers for photo-ionization and cooling.} \label{tab:lasers}
    \end{table*}
    
    \section{Vaporization}
    
\begin{figure}
        \centering
        \includegraphics[width=.45\textwidth]{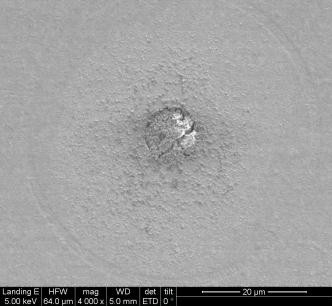}
        \caption{SEM image of the surface of a Sandia ring trap post test with blocked loading hole. Figure courtesy of Sandia National Labs.}
        \label{fig:ba-block}
    \end{figure}

    Obtaining a beam of neutral atoms presents some unique challenges when trapping barium in a chip trap.
    \begin{enumerate}
        \item Barium oxidizes readily in air and consequently the barium source must be protected from air during installation.
        \item The fluorescence of neutral barium atoms at \SI{791}{\nm} is extremely weak and it has not been possible to use it to confirm the presence of neutral barium.
        \item Blocking of the small loading apertures of chip traps with barium clusters can render a trap useless. Figure~\ref{fig:ba-block} shows a SEM image of the surface of a Sandia ring trap after testing with a traditional barium oven. This chip was also imaged with an energy-dispersive X-ray spectrograph (EDS) and the blockage identified as barium oxide.
    \end{enumerate}
    
    The traditional approach to the oxidation problem has been to arrange an inert gas atmosphere and load fresh barium metal flakes into a thin alumina tube which is heated by a tungsten coil. Despite this precaution, some oxide is inevitably formed and the oven must initially run significantly hotter than usual to clean off the oxide layer. It is during this initial `break-in' of a traditional oven that we expect the majority of the clusters seen in Figure~\ref{fig:ba-block} originate.
    
    We have adopted a 2-part solution for reliable generation of neutral barium.
    \begin{enumerate}
        \item We use a commercial barium oven\footnote{Manufactured by Alvatec Alkali Vacuum Technologies GmbH} in which the barium is contained in a stainless-steel tube filled with argon and sealed with indium. The indium seal breaks when the oven is heated. In operation the oven is heated by running current of up to \SI{10}{\ampere} through the tube. A small current (typically \SI{5}{\ampere}) is run through the oven during the bake-out of the chamber.
        \item An oven shutter has been developed. This consists of a thermostatic bimetal strip\footnote{Shivalik Bimetals type 721-112. Metals are 74-24-1 Alloy and Invar 155 Alloy.} cut into an elongated `U' shape with a stainless steel shutter attached. This is mounted to a Macor\textsuperscript{\textregistered} block and mounted above the oven. When activated by running current through the strip it bends and moves the shutter to cover the oven tube. This protects the chip from excess barium flux and any indium from the seal. This is activated when the oven is run initially and continuously during chamber bake-out.
    \end{enumerate}
    
    \section{Electrical setup}

    Because of the large number of DC electrodes which must be controlled, using off-the-shelf equipment quickly becomes cumbersome and expensive.

\begin{figure}
        \centering
        \includegraphics[width=.45\textwidth]{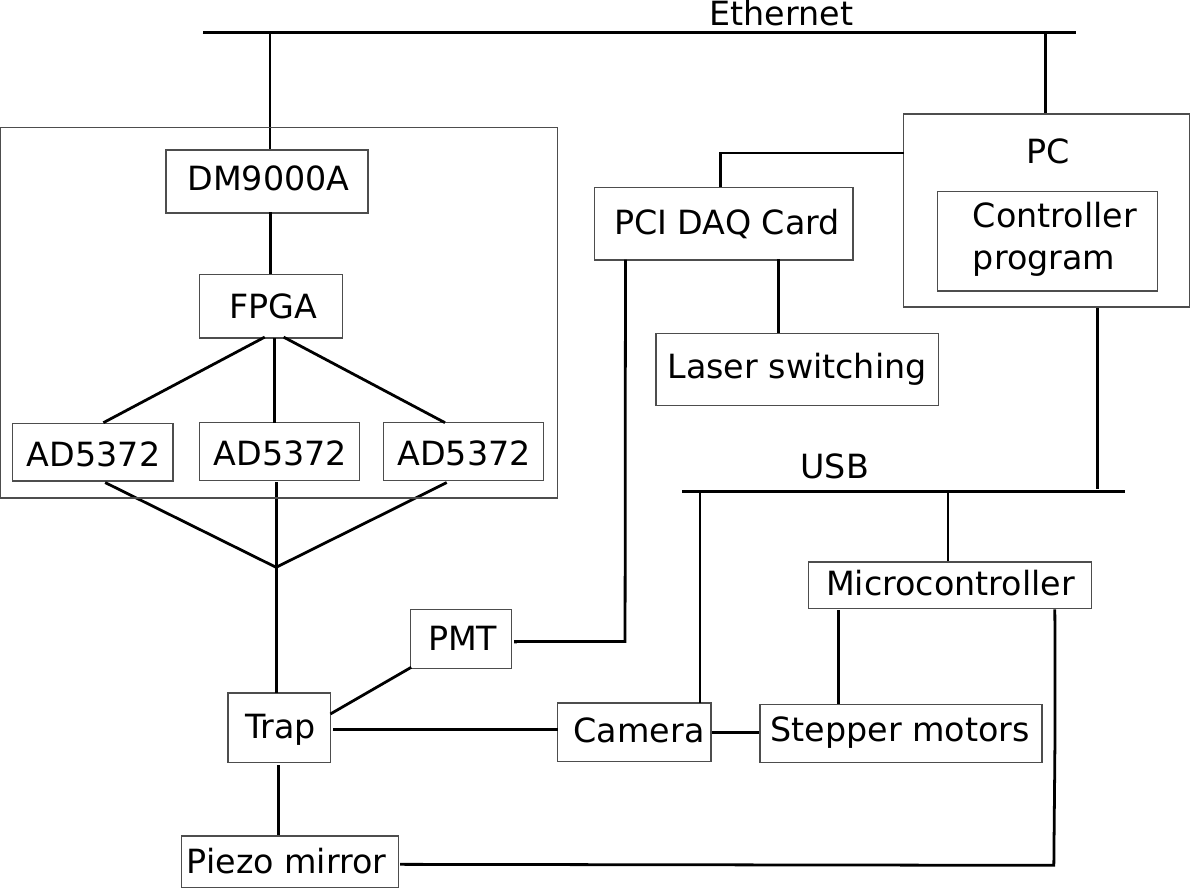}
        \caption{Block diagram of the electrical setup of the experiment control hardware. The PC controls the trap electrodes via the FPGA system. The PC also sends data to a microcontroller which is responsible for moving the lasers and camera to follow the ion. Camera data is sent to the computer via USB and PMT counts are recorded by a PCI DAQ card. This card is also used to control the lasers.}
        \label{fig:daq-system}
    \end{figure}

    We have designed a custom electrode control system. It is based on three AD5372\cite{datasheet.AD5372} digital to analog converters (DACs) and controlled by a Cyclone II 2C35 field-programmable gate array (FPGA)\cite{datasheet.2C35}. A block diagram of the system is shown in Figure~\ref{fig:daq-system}. Electrode control solutions for shuttling ions about the chip are supplied by the manufacturer and are typically calibrated for a calcium ion with voltages in the range of $\pm$\SI{10}{\volt}. A control program running on the host PC parses the files and applies a chip and ion-species specific mapping to determine the voltages and corresponding DAC channels. The full sequence of bytes required to control the DACs is then calculated and transmitted via UDP\footnote{User Datagram Protocol} over an ethernet connection to the DAC system. The FPGA reads the data via a DM9000A ethernet interface and saves it to SRAM. Upon receipt of a control signal, the FPGA will then write the data in sequence to the three DACs.

    This system has proven quite capable, allowing us to control 96 electrodes over a \SI{20}{\volt} range ($\pm 10$V) to within \SI{0.3}{\milli\volt} at an update rate of \SI{2.4}{\micro\second} per channel with the ability to synchronize channel updates. Though rated for a \SI{2.2}{\nano\farad} load, internal DAC output amplifiers are sufficient to drive the in-vacuum electronics at these speeds. The final update rate depends on the number of channels that need to be updated per step; this is typically around 5--10. The system is also low cost (around \$500) and compact. Source code (python control program and Verilog FPGA code) and schematics are available to other researchers on request.
    
    The application of radio frequency (RF) to the trap is conventional. We use a copper can helical resonator to filter and impedance match from a power amplifier to the RF rails of the trap. The internal RF wires are shielded and take the shortest route possible from the feed-through to the trap. The design of helical resonators for this purpose has been explored by Siverns \emph{et.al.}~\cite{siverns.apb.2012.resonators}.
        
    \section{Trap performance}
    
\begin{figure}
        \centering
        \includegraphics[width=.45\textwidth]{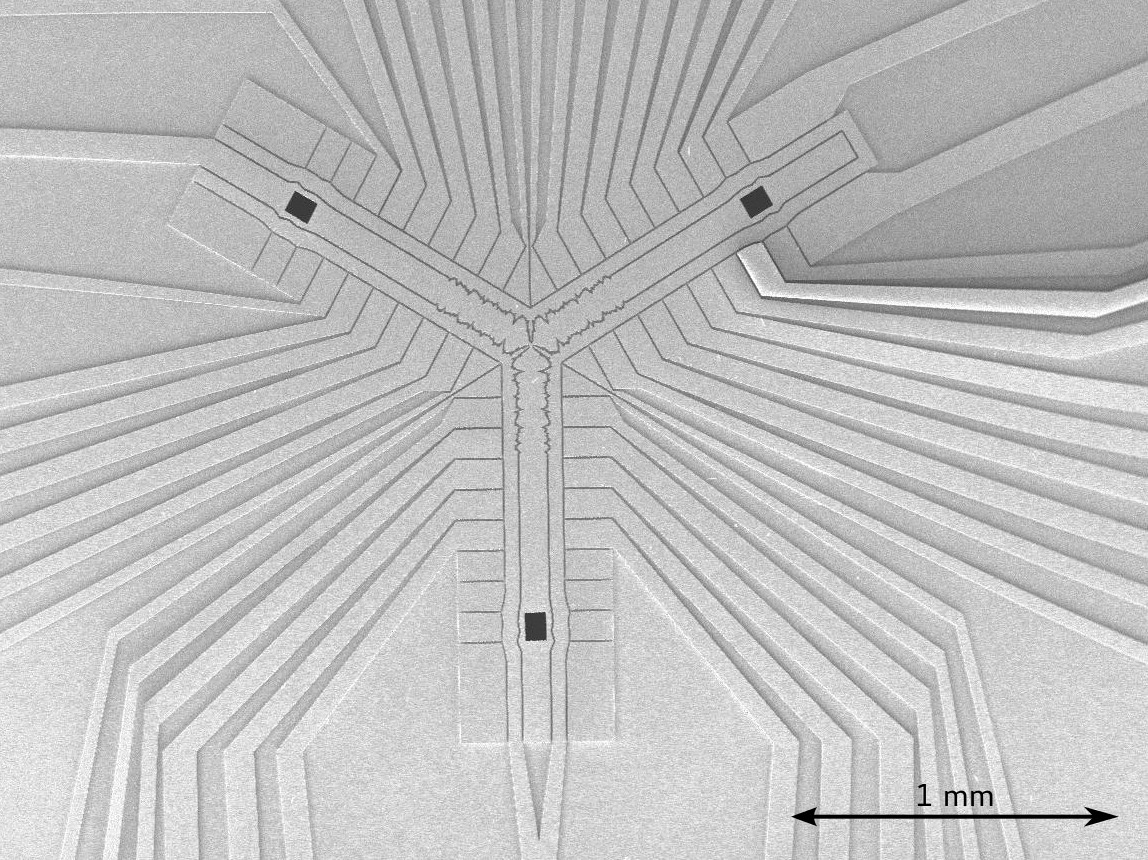}
        \caption{SEM image of the `Y' trap surface. The black rectangular sections are the loading holes. Image courtesy of Sandia National Labs.}
        \label{fig:y-surface}
    \end{figure}

    We have demonstrated trapping of single ions in our system with a `Y' trap manufactured by Sandia National Labs \cite{moehring.njp.2011}. A SEM image of the surface of this trap is shown in Figure~\ref{fig:y-surface}.
    
\begin{figure*}
        \centering
        \begin{subfigure}[b]{0.45\textwidth}
                \centering
                \includegraphics[width=\textwidth]{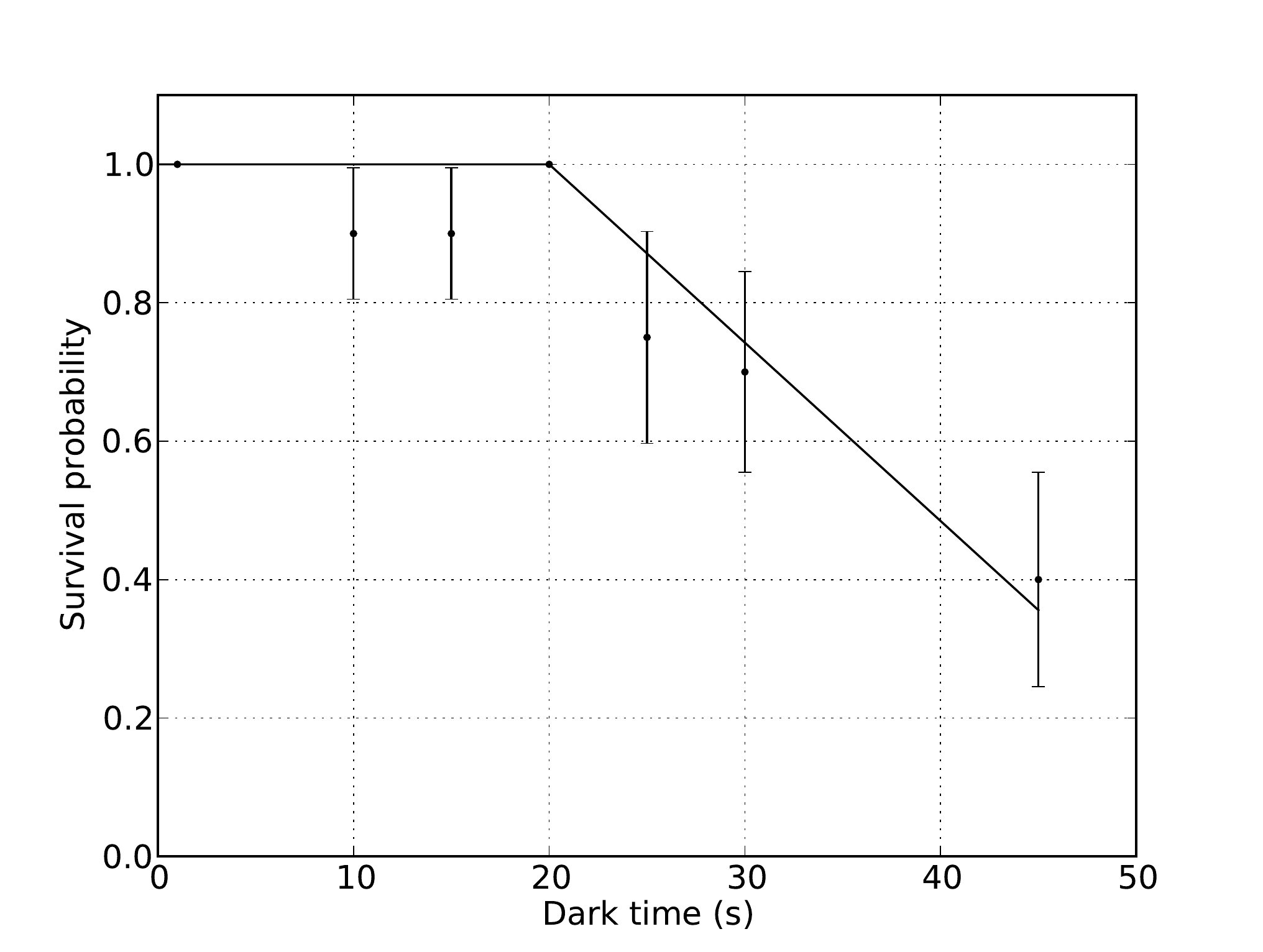}
                \caption{In the loading region}
                \label{fig:lifetime-loading}
        \end{subfigure}%
        ~ 
        \begin{subfigure}[b]{0.45\textwidth}
                \centering
                \includegraphics[width=\textwidth]{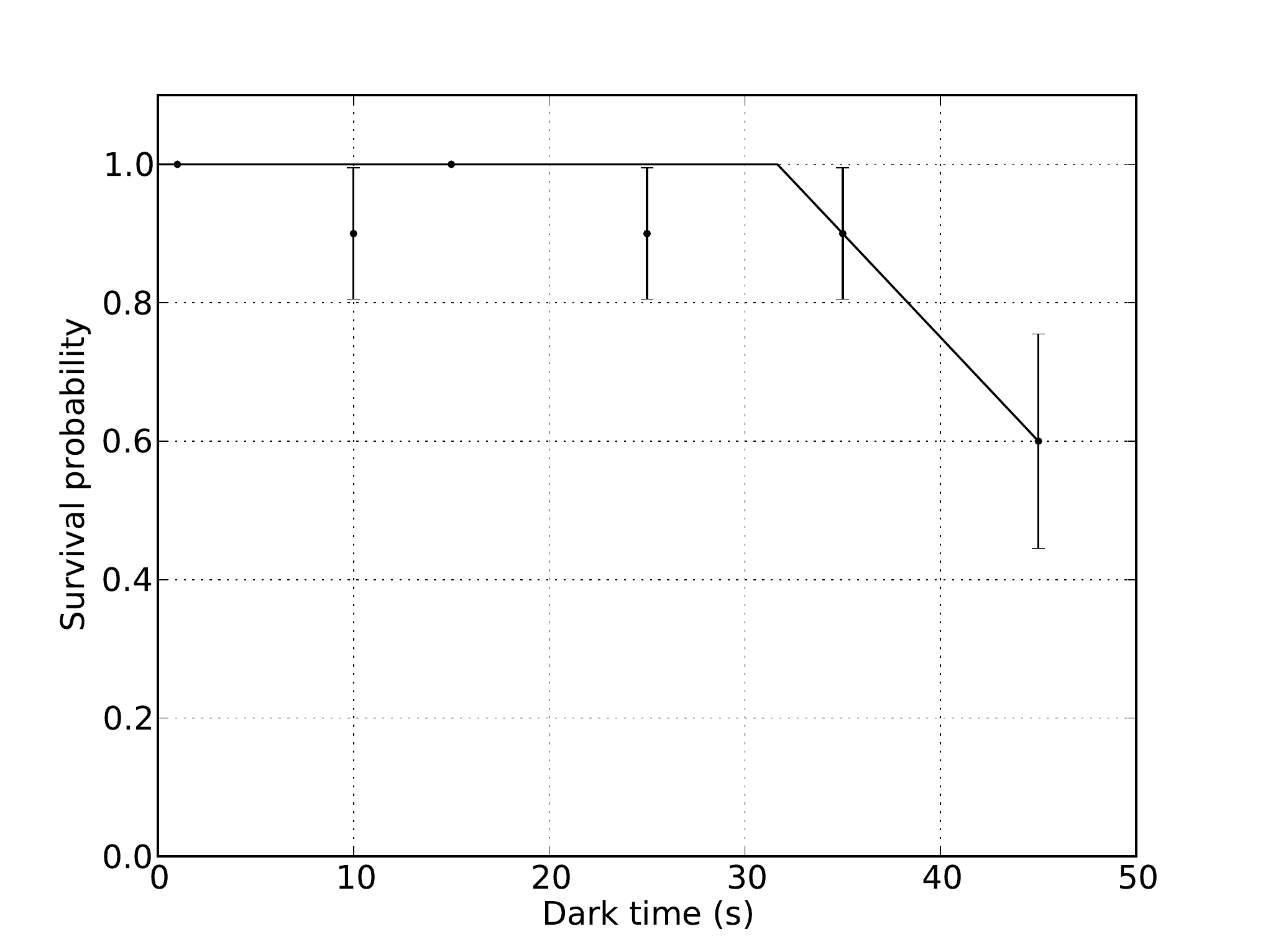}
                \caption{In the middle of a trap arm}
                \label{fig:lifetime-midarm}
        \end{subfigure}
        \caption{Ion dark lifetime as measured following interrupting of the laser cooling. The fit to the data takes the form of Equation~\ref{eqn:lifetime}. Dark lifetimes are \SI{20 \pm 4.9}{\second} and \SI{31.6 \pm 3.4}{\second} for the loading region and mid-arm location respectively.
        }\label{fig:lifetimes}
    \end{figure*}

    Ion dark lifetimes were measured. This was done by repeatedly mechanically shuttering the re-pump laser\footnote{Entirely shuts off the cooling in barium.} for a varying length of time and finding the probability $p_{l}$ of ion loss after cooling was re-established. Dark lifetimes have not been defined in a consistent way in the literature. In practice we are mostly interested in a measurement of the delay time over which all loss mechanisms are insignificant. Therefore we define the dark lifetime $\tau_{d}$ by making a weighted fit of 
    \begin{equation}\label{eqn:lifetime}
        f(t) = \begin{cases}
            t < \tau_d, & 1 \\
            t \geq \tau_d, & 1-a (p_{l}-\tau_d)
        \end{cases}
    \end{equation}
    as shown in Figures~\ref{fig:lifetime-loading} and \ref{fig:lifetime-midarm}. Uncertainties for each point are $\sqrt{p_l(1-p_l)/N}$, found assuming a binomial distribution where $N$ is the number of measurements and the final parameter uncertainties are found from the covariance matrix produced after fitting with the Levenburg-Marquardt algorithm.
    
    A dark lifetime of \linebreak[4]\SI{20 \pm 4.9}{\second} was measured with the ion directly over the loading region. After shuttling the ion to the mid point of one of the trap arms an improved dark lifetime of \SI{31.6 \pm 3.4}{\second} was measured. Some authors prefer the lifetime corresponding to 50\% ion loss, which in our approximation is simply $\frac{1}{2a} + \tau_d$. Our corresponding lifetime values in this formulation are more uncertain, \SI{39 \pm 22}{\second} and \SI{48 \pm 27}{\second} respectively. 
        
    The secular frequencies of the trap were measured by the resonant excitation technique \cite{Sudakov200010}. These were found to be \SI{.75}{\mega\hertz}, \SI{1.5}{\mega\hertz} and \SI{2.1}{\mega\hertz}. Given these we were able to determine a RF voltage of \SI{96}{\volt}. This was done using boundary element model (BEM) solutions \cite{slusher2010scalable} of the surface potentials of this trap \footnote{BEM solutions generously supplied by C.E.\~Volin, Georgia Tech Research Institute.} and the known electrode voltages. These were combined with the RF pseudo-potential calculated for a range of realistic RF voltages (around 50 --- 150 V) to form secular potential solutions at each of which the secular frequencies were solved and the best match found.
    
    The corresponding classical well depth was found by slicing the calculated potential through the weak axis at the trapping location. We find a classical well depth of \SI{44}{\milli\eV}. Note that this technique does not take into account the strong stray fields above the loading region or any additional stray surface fields which are typical of these aluminum surface traps. Consequently it is most useful as a representative figure and as a point of comparison between chips of similar design.
    
    We have repeated the calculation of the well depth at regions outside of the loading region and find comparable depths; around \SI{48}{\milli\eV}. Therefore, we expect the significant difference in lifetime between locations is due to excess stray fields in the loading region rather than the different trap geometry.
    
    \section{Conclusion}
    
    After operating two systems of this design for nearly two years we have found them reliable and easy to assemble. Traps can be swapped out and modifications made in a few hours; a process which has been done at least 10 times. Operating pressure (around $10^{-11}$~Torr) is readily achieved after a few days of baking followed by titanium sublimation pumping. Despite the unconventional design of the system we see no evidence of any associated degradation in trap performance. The signals applied to the DC electrodes are clean, owing to the isolation from the in-vacuum filters. The AD5372 based control solution provides sufficient drive to allow convenient control of the DC voltages and is suitable for setting up trapping regions and shuttling operations.
    
    \begin{acknowledgments}
    This research is supported by Intelligence Advanced Research Projects Activity group (IARPA).\\
    We also acknowledge Rogers Corporation and Mill-Max Mfg.\ Corp for proving free sample materials.
    \end{acknowledgments}

\clearpage
\bibliography{chiptrap_references}

\end{document}